\documentclass[11pt,a4paper]{article}
\usepackage{amsmath,amssymb,amsfonts}
\usepackage{multirow}
\usepackage{jheppub}
\IfFileExists{orcidlink.sty}{\usepackage{orcidlink}}{%
  \providecommand{\orcidlink}[1]{}%
}
\usepackage{bm}
\usepackage{comment}
\usepackage{booktabs}
\usepackage{array}
\usepackage{units}
\usepackage{xcolor,slashed}
\usepackage{xspace}

\hypersetup{
  colorlinks=true,
  linkcolor=blue!50!black,
  citecolor=blue!50!black,
  urlcolor=blue!50!black
}

\preprint{DESY-26-110}

\newcommand{\tev}{\mathrm{TeV}}
\newcommand{\gev}{\mathrm{GeV}}
\newcommand{\wilson}{\mathcal{C}}
\newcommand{\order}{\mathcal{O}}

\newcommand{\dalphaHadMZ}{\ensuremath{\Delta\alpha_{\rm had}^{(5)}(M_Z)}\xspace}
\newcommand{\asZ}{\ensuremath{\alpha_{s}(M_Z)}\xspace}

\let\oldtable\table
\let\endoldtable\endtable
\renewenvironment{table}[1][]{%
  \oldtable[#1]%
  \setlength{\abovecaptionskip}{4pt}%
  \setlength{\belowcaptionskip}{4pt}%
}{%
  \endoldtable
}

\title{Electroweak and Single Top-Quark Conspiracy in Current LHC Data}
\author[a]{Christoph Englert\orcidlink{0000-0003-2201-0667},}
\author[b]{Roman Kogler\orcidlink{0000-0002-5336-4399},}
\author[c]{and Michael Spannowsky\orcidlink{0000-0002-8362-0576}}
\affiliation[a]{Department of Physics \& Astronomy, University of Manchester, Manchester M13 9PL,\\United Kingdom}
\affiliation[b]{Deutsches Elektronen-Synchrotron DESY, Notkestr. 85, 22607 Hamburg, Germany}
\affiliation[c]{Institute for Theoretical Physics, Campus S\"ud, Karlsruhe Institute of Technology (KIT), D-76128 Karlsruhe, Germany}
\emailAdd{christoph.englert@manchester.ac.uk}
\emailAdd{roman.kogler@desy.de}
\emailAdd{michael.spannowsky@kit.edu}
\abstract{
Recent LHC measurements of electroweak single top-quark and top-quark associated Higgs boson production show a number of rate shifts, while related multiboson channels remain close to their Standard Model predictions. Gauge symmetry relates these processes such that deviations can be organised as the low-energy imprint of a common ultraviolet origin. A Standard Model Effective Field Theory (SMEFT) analysis of these rates together with $Z$-pole observables identifies the pattern: a weak top dipole and a triple-gauge deformation carry the largest pulls, accompanied by a positive left-handed top-quark current. An acceptable fit is also obtained with the pure triple-gauge coefficient set to zero. We then ask whether a gauge-invariant model can produce this set of operators and still pass the numerous constraints from experimental data. Within weakly coupled renormalisable matching, no single scalar, vector or vector-like-quark multiplet is sufficient to reproduce the full pattern. A minimal possibility is a multi-threshold vector-like-quark sector (a singlet, a doublet and two triplets) whose currents arise at tree level, while the dipole requires additional loop dynamics that the current sector does not fix. A representative realisation, confronted with direct, indirect and flavour constraints, points to correlated measurements of electroweak single top-quark production, improved $tH$ and $t \bar t H$ sensitivity, and updated searches for vector-like quarks.
}

\begin{document}
\maketitle
\allowdisplaybreaks
\flushbottom
\section{Introduction}
With no direct evidence for new heavy particles at the LHC, the first signs of a new sector may instead appear as moderate shifts across processes governed by the same electroweak, top quark and Higgs interactions. We ask whether the present measurements can be described by a common set of gauge-invariant interactions without spoiling the agreement elsewhere.

Unlike inclusive top quark pair production, which is dominated by QCD, the electroweak processes $tq\gamma$, $tWZ$ and $tH$ probe charged-current top couplings, electroweak dipoles and Higgs--gauge interactions already at leading order. Gauge invariance ties these interactions to the $Wtb$, $Zt\bar t$ and $Zb\bar b$ vertices, and also to vector-boson scattering, triboson production and Higgs boson observables. Any common explanation of the top-associated rates must therefore also be compatible with the multiboson and precision measurements.

The pattern considered here is set by the recent $tWZ$, $tq\gamma$ and top-quark associated Higgs boson measurements. The $tWZ$ and $tq\gamma$ central values lie above the SM predictions~\cite{CMS:2025tre,ATLAS:2023qdu}, while the ATLAS multilepton analysis finds a large, though imprecise, $tH$ contribution together with a $t\bar tH$ rate below the SM expectation~\cite{ATLAS:2025eua}. Vector-boson scattering and triboson production probe the same interactions but remain broadly compatible with the SM at the present precision~\cite{CMS:2026add,ATLAS:2024nab,CMS:2025hlu}. We use these channels to test whether the shifts in the top-associated rates can be accommodated without producing comparable effects in the multiboson sector.

Electroweak precision observables (EWPOs) provide a direct consistency test of an interpretation based on a modified left-handed current. Since $Q_3=(t_L,b_L)$, a change in the left-handed top current is tied by gauge invariance to the $Zb_L\bar b_L$ coupling. We therefore include five LEP and SLC pseudo-observables in the fit, including the long-standing tension in the $b$-quark forward--backward asymmetry. The detailed inputs and their treatment are given in Section~\ref{sec:response-map}.

To test these correlations in a common parametrisation, we perform a linear fit to five SMEFT coefficients~\cite{Grzadkowski:2010es,Brivio:2020onw}. The operator set separates a pure triple-gauge interaction, a Higgs--gauge interaction, the weak top dipole and the left-handed third-generation current. A second current coefficient is included because the $Zb_L\bar b_L$ data constrain a combination of the two currents rather than the left-handed charged-current coefficient alone. The weak dipole has the largest pull from zero, while the left-handed current has a positive best-fit value. The Higgs--gauge coefficient is poorly determined. The fit remains acceptable when the pure triple-gauge coefficient is set to zero, so the matching discussion concentrates on the current and dipole coefficients.

The fit identifies the interactions that need to be matched, but not whether they have a consistent ultraviolet origin. We therefore ask whether a gauge-invariant heavy sector can generate the current and dipole coefficients without violating electroweak precision, Higgs, flavour or direct-search bounds. At leading order in weakly coupled renormalisable matching, no one-particle simplified model suffices. A scalar does not generate the protected third-generation current structure at tree level; a vector does not generate the weak dipole at tree level, and its gauge completion brings additional states; and, within our heavy--light Yukawa ansatz, a single vector-like-quark multiplet cannot supply the two chiral mixings. The positive current further requires custodial partners to protect $Zb_L\bar b_L$. The construction is therefore intrinsically multi-field, with the dipole generated by further interactions among the heavy states.

\begin{figure}[tb]
\centering
\includegraphics[width=0.8\textwidth]{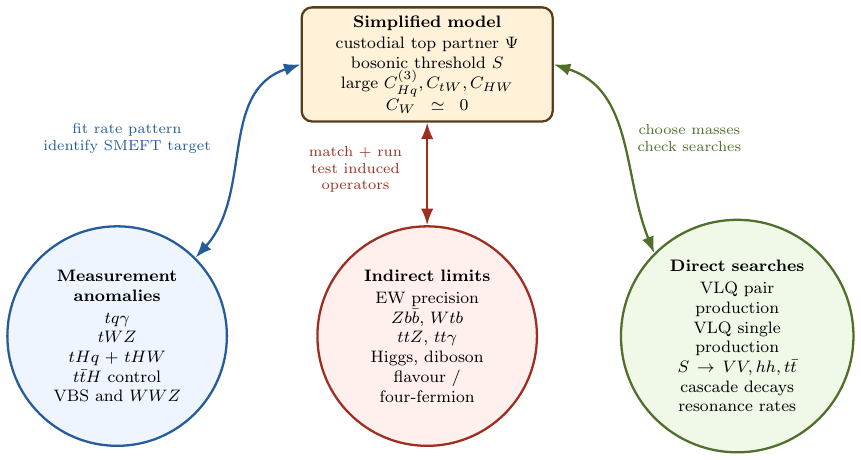}
\caption{Outline of the analysis. Electroweak top quark, Higgs boson, multiboson and precision measurements constrain the SMEFT coefficients. Tree-level matching relates the current operators to a gauge-invariant vector-like-quark sector, while the dipole requires additional heavy-sector interactions. The resulting spectrum and couplings are compared with indirect constraints and published direct-search limits.}
\label{fig:narrative}
\end{figure}

Accordingly, we use tree-level matching of the current coefficients to guide the choice of fermion representations and their custodial alignment, while parametrising the additional one-loop dipole contribution by an effective form factor. An aligned benchmark then allows us to follow the correlated consequences for Higgs boson and top-quark couplings, flavour observables and direct searches. These constraints cannot be treated one at a time: a change made to satisfy one bound generally changes the matching or the predictions elsewhere. An overview of the analysis is shown in Figure~\ref{fig:narrative}. 

Since our aim is to test this particular operator pattern rather than reproduce a global SMEFT analysis, the likelihood uses rate information from the selected channels. Section~\ref{sec:response-map} gives the input mappings and the approximations used in the fit. For global SMEFT analyses including these sectors, see Refs.~\cite{Ellis:2018gqa,Biekotter:2018ohn,Madigan:2022cvc}.

The paper is organised as follows. Section~\ref{sec:response-map} defines the operator set and likelihood and presents the fit. Section~\ref{sec:thresholds} develops the gauge-invariant heavy-fermion sector and its custodially aligned benchmark. We then examine the correlated Higgs, top and flavour effects and compare the spectrum with published direct-search limits before summarising our conclusions.

\section{Five-parameter SMEFT fit}
\label{sec:response-map}
\subsection{Operator basis}

\begin{figure}[!b]
\centering
\begin{minipage}[t]{0.34\textwidth}
\centering
\includegraphics[width=0.82\linewidth]{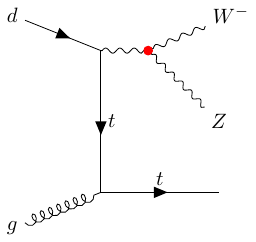}\\[-1mm]
{\footnotesize $\order_W$ in $tWZ$}
\end{minipage}
\vspace{0.4cm}
\begin{minipage}[t]{0.34\textwidth}
\centering
\vspace{-3.39cm}
\includegraphics[width=0.84\linewidth]{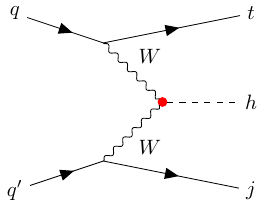}\\[-1mm]
{\footnotesize $\order_{HW}$ in $tHq$}
\end{minipage}
%
\includegraphics[width=0.68\textwidth]{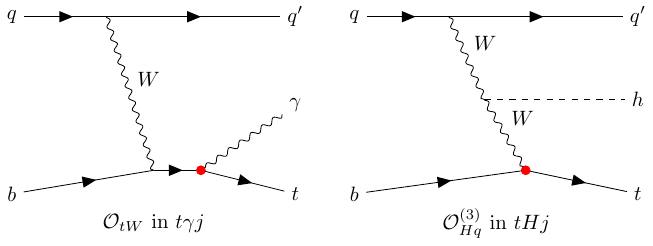}
\caption{Representative topologies involving the four operators that enter the LHC rates. The red dot denotes the effective interaction insertion. Top left: $\order_W$ in $tWZ$ production. Top right: $\order_{HW}$ in Higgs radiation. Bottom left: $\order_{tW}$ in $t\gamma j$ production. Bottom right: $\order_{Hq}^{(3)}$ in $tHq$ production. These examples are illustrative; each measurement generally receives further SMEFT contributions.} \label{fig:operator-diagrams}
\end{figure}

To identify the interactions that can be shared by the top-quark associated, Higgs boson, multiboson and $Z$-pole measurements, we use a five-coefficient, CP-even Warsaw-basis fit. The retained operators connect these channels most directly. Four-fermion operators and light-quark currents can be important in energetic tails, while flavour-changing top operators describe a production mechanism distinct from the SM-like $t$-channel topology used here for $tq\gamma$. CP-odd interactions require observables beyond the inclusive rates considered in this study. Figure~\ref{fig:operator-diagrams} shows representative insertions of the four operators that enter the LHC rates.

The remaining Wilson coefficients are held fixed so that the fit isolates the charged-current, dipole and bosonic structures probed jointly by these measurements. The omitted set includes the top-Yukawa interaction $\order_{uH}$, which correlates $tH$ and $t\bar tH$, and universal Higgs and oblique operators such as $\order_{H\Box}$, $\order_{HD}$ and $\order_{HWB}$, which are better constrained in a comprehensive Higgs and electroweak fit. Neutral-current top interactions and the hypercharge and chromomagnetic dipoles, including $\order_{Hu}$, $\order_{tB}$ and $\order_{tG}$, are important correlated effects in ultraviolet models, but they add no independent charged-current structure to the fit considered here. We return to several of them in the heavy-sector discussion.

Within this restricted basis, four independent effects must be separated in the collider rates. The left-handed current $\order_{Hq}^{(3)}$ contributes to $t\gamma j$, $tWZ$ and $tHq$, while the weak dipole $\order_{tW}$ modifies electroweak radiation from the top line rather than producing a universal rescaling. The Higgs--gauge operator $\order_{HW}$ links Higgs radiation with the multiboson channels, and $\order_W$ supplies an independent triple-gauge interaction. A fifth operator, $\order_{Hq}^{(1)}$, is needed not to add another LHC rate deformation but to treat the precision constraint correctly: the tree-level $Zb_L\bar b_L$ coupling depends on $C_{Hq}^{(1)}+C_{Hq}^{(3)}$. Its contribution to the LHC rates is neglected in the numerical fit, while its EWPO dependence is retained. The five operators are
\begin{align}
  \order_W &=
    \epsilon^{abc}
    W^{a\,\mu}{}_{\nu}
    W^{b\,\nu}{}_{\rho}
    W^{c\,\rho}{}_{\mu},\\
  \order_{HW} &=
    H^\dagger H\, W^a_{\mu\nu} W^{a\,\mu\nu},\\
  \order_{tW} &=
    (\bar Q_3^{\,i}\sigma^{\mu\nu}t_R)
    (\tau^a)_i{}^j\,\tilde H_j\,W^a_{\mu\nu}
    + {\rm h.c.},\\
    \order_{Hq}^{(1)} &=
    (H^\dagger i\overleftrightarrow{D}_{\mu} H)
    (\bar Q_3 \gamma^\mu Q_3), \\  
  \order_{Hq}^{(3)} &=
    (H^\dagger i\overleftrightarrow{D}^{\,a}_{\mu} H)
    (\bar Q_3^{\,i}(\tau^a)_i{}^j\gamma^\mu Q_{3j}),
\end{align}
where $Q_3$ denotes the left-handed third-generation quark doublet,
$i,j$ are $SU(2)_L$ indices, and $\tau^a$ are the Pauli matrices (not
$\tau^a/2$); repeated $SU(2)_L$ indices are summed. If the dimensionless
Warsaw coefficients are denoted by $\wilson_a$, we use
$C_a\equiv\wilson_a/\Lambda^2$ throughout and quote this dimensionful
combination in $\tev^{-2}$. Thus
\begin{equation}
  {\cal L}_{\rm SMEFT}={\cal L}_{\rm SM}+\sum_a C_a\order_a.
\end{equation}
With this assignment, the LHC likelihood depends on $C_W$, $C_{HW}$, $C_{Hq}^{(3)}$ and $C_{tW}$, whereas $C_{Hq}^{(1)}$ is constrained through the electroweak observables. We profile $C_{Hq}^{(1)}$ but omit it from the main results.

The collider fit is defined directly with fixed Wilson coefficients: RG evolution between matching and process scales is neglected there and included only in the flavour check below.

\subsection{Fit setup}
To test whether coefficient shifts favoured by the top-quark associated channels remain compatible with multiboson and precision data, we construct a common likelihood. It combines the eight LHC rate inputs in Table~\ref{tab:measurements} with the five electroweak precision observables in Table~\ref{tab:ewpo}. For the collider measurements, we identify each input with the parton-level process shown in the second column. This provides a common SMEFT parametrisation, but it does not reconstruct detector efficiencies, EFT-dependent acceptances or the full experimental covariance matrices.

Putting the two single top-quark measurements on a common rate footing requires slightly different treatments. The CMS $tWZ$ result combines data at $13$ and $13.6~\tev$ and reports a signal strength of $1.77\pm0.32$~\cite{CMS:2025tre}. For $tq\gamma$ we form the signal strength from the ATLAS fiducial measurement
\begin{equation}
  \sigma(tq\gamma)\,B(t\to \ell\nu b)
  = 688 \pm 23\,{}^{+75}_{-71}\,\unit{fb},
\end{equation}
and the NLO SM prediction $515^{+36}_{-42}$\,\unit{fb}~\cite{ATLAS:2023qdu}. Their ratio gives the signal strength quoted in Table~\ref{tab:measurements}.

Because the published rate inputs do not all correspond one-to-one to a single parton-level calculation, we make the following assignments. The $tq\gamma$ measurement is represented by $pp\to t\gamma j$ and its charge conjugate rather than by a fiducial recast. The ATLAS $tH$ measurement of $7.2^{+4.6}_{-4.0}$ is the $tHq$ component of a joint analysis with the $t\bar tH$ signal strength extracted from the same simultaneous fit, where we include the quoted correlation of $-0.11$ between the two. In this measurement, the small $tHW$ component is set to the SM value. 
Both triboson inputs are taken with resonant $VH$ production treated as background. In this case, the signal definition of the ATLAS $VVZ$ measurement includes $WZZ$ and $ZZZ$ admixtures of about 30\%, which we evaluate as pure $WWZ$.
Finally, the $tWZ$ measurement combines two collision energies while its SMEFT dependence is evaluated with a single parton-level parametrisation.

The $Z$-pole inputs test the same left-handed current sector through the combination $C_{Hq}^{(1)}+C_{Hq}^{(3)}$. 
The total $Z$-boson width is taken from the LEP and SLC combination~\cite{ALEPH:2005ema}, $\Gamma_Z = 2.4955 \pm 0.0023~\gev$, updated for the improved Bhabha cross section~\cite{Janot:2019oyi}. 
The ratios of partial $Z$-boson decay widths, $R_\ell^0 \equiv \Gamma_{\rm had}/\Gamma_{\ell\ell}= 20.767 \pm 0.025$ and $R_b^0 \equiv \Gamma_{b\bar b}/\Gamma_{\rm had}= 0.21629 \pm 0.00066$ and the $b$ quark forward-backward asymmetry $A_{\rm FB}^{0,b} = 0.0996 \pm 0.0016$ have been measured at LEP~\cite{ALEPH:2005ema}, where $A_{\rm FB}^{0,b}$ has been corrected for the $m_b$ dependence~\cite{Bernreuther:2016ccf} of the two-loop QCD correction~\cite{Catani:1999nf,Djouadi:1989uk}. 
The longitudinal beam polarisation at SLC allowed the SLD Collaboration to determine the $b$-quark asymmetry parameter directly, $A_b = 0.923 \pm 0.020$~\cite{Abe:2000dq,Abe:2000uc,Abe:2000hk}. Out of these five observables, only $A_{\rm FB}^{0,b}$ shows a significant deviation from the SM prediction, which amounts to 2.3 standard deviations~\cite{Fischer:2026bka, ParticleDataGroup:2026}.

\begin{table}[t]
\centering
\caption{LHC rate measurements included in the fit and the parton-level processes used to calculate their SMEFT dependence. The uncertainties are implemented as shown and remain asymmetric where applicable; theory components are not treated uniformly across the source measurements.}
\label{tab:measurements}
\begin{tabular}{llc}
\toprule
Measurement & Parton-level process & $\mu_{\rm obs}$ \\
\midrule
ATLAS $tq\gamma$~\cite{ATLAS:2023qdu} & $t\gamma j$ & $1.34^{+0.19}_{-0.17}$ \\
CMS $tWZ$~\cite{CMS:2025tre} & $tWZ$ & $1.77\pm0.32$ \\
ATLAS $tHq$~\cite{ATLAS:2025eua} & $tHq+tHW$ & $7.2^{+4.6}_{-4.0}$ \\
ATLAS $t\bar tH$~\cite{ATLAS:2025eua} & $t\bar tH$ & $0.59^{+0.22}_{-0.20}$ \\
CMS VBS $W^\pm W^\pm$~\cite{CMS:2026add} & $ssWWjj$ & $1.05^{+0.15}_{-0.14}$ \\
CMS VBS $WZ$~\cite{CMS:2026add} & $WZjj$ & $1.21^{+0.28}_{-0.25}$ \\
ATLAS $VVZ$~\cite{ATLAS:2024nab} & $WWZ$ & $1.59\pm0.38$ \\
CMS $WWZ$~\cite{CMS:2025hlu} & $WWZ$ & $0.87\pm0.45$ \\
\bottomrule
\end{tabular}
\vspace{2.5ex}
\caption{Electroweak precision observables included in the fit. The third column gives the corresponding predictions from the global electroweak analysis of Ref.~\cite{Fischer:2026bka}. The pull is the difference between measurement and prediction divided by the measurement uncertainty. The fit uses these five observables without the complete LEP/SLD covariance matrix.}
\label{tab:ewpo}
\begin{tabular}{lccc}
\toprule
Pseudo-observable & Measurement & Reference prediction & Pull \\
\midrule
$\Gamma_Z$~[GeV]   & $2.4955 \pm 0.0023$   & $2.4945 \pm 0.0006$   & $+0.4$ \\
$R_\ell^0$         & $20.767 \pm 0.025$    & $20.750 \pm 0.008$    & $+0.7$ \\
$R_b^0$            & $0.21629 \pm 0.00066$ & $0.21589 \pm 0.00009$ & $+0.6$ \\
$A_{\rm FB}^{0,b}$ & $0.0996 \pm 0.0016$   & $0.1032 \pm 0.0003$   & $-2.3$ \\
$A_b$              & $0.923 \pm 0.020$     & $0.93475 \pm 0.00004$ & $-0.6$ \\
\bottomrule
\end{tabular}
\end{table}

To treat the collider and precision inputs at the same order in the effective expansion, we retain the interference between the SM and dimension-six amplitudes at $\mathcal O(\Lambda^{-2})$. For each signal strength $\mu_i$, this gives
\begin{equation}
  \mu_i(C_a)=
  1+\sum_a \ell_{ia}C_a,
  \label{eq:rate_model}
\end{equation}
where the slopes $\ell_{ia}$ are extracted from scans at $C_a=\pm1~\tev^{-2}$ simulated with \texttt{SMEFTsim}~\cite{Brivio:2017btx, Brivio:2020onw} interfaced with \texttt{MadGraph}~\cite{Alwall:2014hca}. The electroweak precision observables are parametrised in the same way,
\begin{equation}
  P_i(C_a)=
  P_i^{\mathrm{SM}} + \sum_a p_{ia}C_a,
  \label{eq:ewpo_model}
\end{equation}
where the coefficients $p_{ia}$ are taken from the next-to-leading-order calculation of Ref.~\cite{Dawson:2019clf}. For third-generation quarks we use the flavour-general results of Refs.~\cite{Dawson:2022bxd,Bellafronte:2023amz}. The SM predictions retain their dependence on the input parameters in the on-shell renormalisation scheme through semi-analytical parametrisations of the full electroweak next-to-next-to-leading-order calculation~\cite{Dubovyk:2019szj}. The SM inputs, their uncertainties and the theory uncertainties follow Ref.~\cite{Fischer:2026bka}.

This linearisation defines the domain in which the numerical results should be read. We truncate all observables at $\mathcal O(\Lambda^{-2})$ and assign no EFT truncation uncertainty. The broad outer intervals, particularly for $C_{HW}$, can extend into regions where dimension-six-squared and dimension-eight contributions need not be subleading. They should therefore be interpreted within the linear fit defined here.

With the collider and precision responses specified, we profile the Wilson coefficients jointly with the SM inputs using the Gfitter framework~\cite{Flacher:2008zq,Baak:2011ze,Baak:2012kk,Baak:2014ora,Haller:2018nnx,Fischer:2026bka}. In addition to the thirteen observables in Tables~\ref{tab:measurements} and~\ref{tab:ewpo}, the fit includes the SM parameters $M_H$, $M_Z$, $m_t$, $\asZ$ and $\dalphaHadMZ$: the Higgs- and $Z$-boson masses, the top-quark mass, the strong coupling, and the five-flavour hadronic contribution to the running electromagnetic coupling at $M_Z$. Experimental uncertainties are represented by independent Gaussian terms, with separate upper and lower widths where available. We do not include correlations of experimental results from systematic uncertainties, except for the one quoted between the ATLAS $tHq$ and $t\bar tH$ results. Responses involving several partonic processes are combined with SM cross-section weights. This likelihood definition is used for all results below.

\subsection{Fit results}

\begin{figure}[tb]
    \centering
    \includegraphics[width=0.65\textwidth]{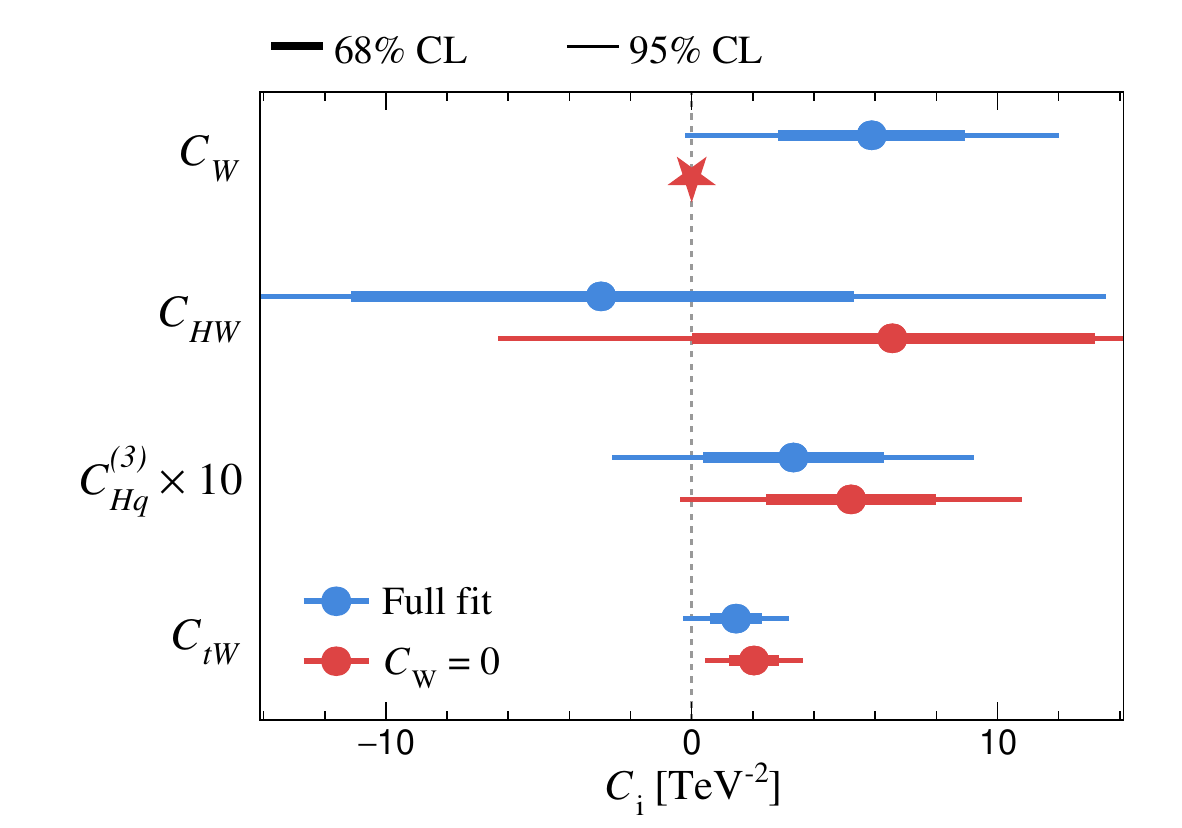}
    \caption{Profile-likelihood results for the eight LHC rate inputs in Table~\ref{tab:measurements} and the five precision observables in Table~\ref{tab:ewpo}. Blue denotes the five-coefficient fit and red the fit with $C_W=0$, where the red star denotes the fixed value of $C_W=0$. The markers show the best-fit values; thick and thin bars give the $68\%$ and $95\%$ profile intervals. Coefficients are quoted in $\tev^{-2}$, the dashed line marks the SM, and $C_{Hq}^{(3)}$ is rescaled by a factor of ten. The coefficient $C_{Hq}^{(1)}$ is profiled in both fits but is not shown.\label{fig:fit_result}}
\end{figure}

Before interpreting individual coefficients, we first ask whether the five-parameter model gives a viable common description of the selected measurements. It does, with a minimum negative log-likelihood value $-2 \ln L = \chi^2_{\min}=11.2$ for eight degrees of freedom (d.o.f.) and a corresponding $p$-value of 0.19. The Wilson coefficients are varied without external constraints, while the SM inputs and theory uncertainties retain their standard constraints and remain within their input ranges at the minimum. Among the four displayed coefficients, $C_W$ has the largest pull from zero, at about $1.9\sigma$, followed by the weak dipole $C_{tW}$ at about $1.7\sigma$. The $95\%$ intervals of both nonetheless still contain zero. The best-fit value of $C_{Hq}^{(3)}$ is positive but less pronounced, whereas $C_{HW}$ is weakly constrained, with a $95\%$ interval of $[-19,14]~\tev^{-2}$. Figure~\ref{fig:fit_result} shows the profile likelihoods, and Table~\ref{tab:gstc_fullfit} gives the corresponding intervals.

\begin{table}
\centering
\caption{The $68\%$ and $95\%$ profile intervals for the four displayed Wilson coefficients. They follow from the operator basis, process assignments and independent-error treatment specified above.
\label{tab:gstc_fullfit}}
\begin{tabular}{lcc}
\toprule
Wilson coefficient & 68\% CL interval $[\tev^{-2}]$ & 95\% CL interval $[\tev^{-2}]$ \\
\midrule
$C_{W}$ & $[2.8, 8.9]$ & $[-0.23, 12]$\\
$C_{HW}$ & $[-11, 5.3]$ & $[-19, 14]$\\
$C_{Hq}^{(3)}$ & $[0.037, 0.63]$ & $[-0.26, 0.92]$\\
$C_{tW}$ & $[0.59, 2.3]$ & $[-0.27, 3.2]$\\
\bottomrule
\end{tabular}
\end{table}

To distinguish the absolute goodness of fit from the improvement obtained by adding the SMEFT coefficients, we also evaluate the same likelihood at the SM point. There, all Wilson coefficients vanish and $\mu_i=1$ for the LHC rates. The precision predictions vary only through the SM inputs. The corresponding minimum is $\chi^2_{\min}=24.6$ for 13 d.o.f., giving $p=0.026$. Allowing all five Wilson coefficients lowers the minimum by
\begin{equation}
  \Delta\chi^2=\chi^2_{\rm SM}-\chi^2_{\rm SMEFT}=24.6-11.2=13.4.
\end{equation}
For five additional coefficients, Wilks' theorem~\cite{Wilks:1938dza} gives $p\simeq0.02$, or a corresponding one-sided significance of about $2.1\sigma$. We note that this is a local significance obtained after a preselection of measurements to study the observed pattern, and it therefore cannot be interpreted as a global improvement over the SM.

\begin{table}[b]
\centering
\caption{Predictions at the best-fit points for all
observables entering the fit (Tables~\ref{tab:measurements} and~\ref{tab:ewpo}).
The pull is the difference between the measurement and the best-fit prediction,
divided by the measurement uncertainty. For an asymmetric measurement, the
displayed pull uses the symmetrised uncertainty $\bar\sigma=(\sigma_++\sigma_-)/2$.
}
\label{tab:bestfit_pred}
\begin{tabular}{lccccc}
\toprule
\multirow{2}{*}{Observable} & \multirow{2}{*}{Measurement} & \multicolumn{2}{c}{Full scan} & \multicolumn{2}{c}{$C_W=0$ scan} \\
\cmidrule(lr){3-4}\cmidrule(lr){5-6}
                            &                              & Prediction & Pull  & Prediction & Pull \\
\midrule
ATLAS $tq\gamma$        & $1.34^{+0.19}_{-0.17}$ & $1.05$  & $+1.6$  & $1.02$ & $+1.7$ \\
CMS $tWZ$               & $1.77\pm0.32$          & $1.69$  & $+0.2$  & $1.32$ & $+1.4$ \\
ATLAS $tHq$             & $7.2^{+4.6}_{-4.0}$    & $0.00$  & $+1.7$  & $1.61$ & $+1.3$ \\
ATLAS $t\bar tH$        & $0.59^{+0.22}_{-0.20}$ & $0.96$  & $-1.8$  & $1.00$ & $-2.0$ \\
CMS VBS $W^\pm W^\pm$   & $1.05^{+0.15}_{-0.14}$ & $1.11$  & $-0.4$  & $1.14$ & $-0.6$ \\
CMS VBS $WZ$            & $1.21^{+0.28}_{-0.25}$ & $1.10$  & $+0.4$  & $1.01$ & $+0.8$ \\
ATLAS $VVZ$             & $1.59\pm0.38$          & $1.43$  & $+0.4$  & $1.05$ & $+1.4$ \\
CMS $WWZ$               & $0.87\pm0.45$          & $1.43$  & $-1.2$  & $1.05$ & $-0.4$ \\
\midrule
$\Gamma_Z$~[GeV]   & $2.4955\pm0.0023$   & $2.4962$  & $-0.3$ & $2.4971$  & $-0.7$ \\
$R_\ell^0$         & $20.767\pm0.025$    & $20.754$  & $+0.5$ & $20.752$  & $+0.6$ \\
$R_b^0$            & $0.21629\pm0.00066$ & $0.21641$ & $-0.2$ & $0.21631$ & $\phantom{+}0.0$ \\
$A_{\rm FB}^{0,b}$ & $0.0996\pm0.0016$   & $0.0997$  & $-0.1$ & $0.0992$  & $+0.2$ \\
$A_b$              & $0.923\pm0.020$     & $0.9346$  & $-0.6$ & $0.9345$  & $-0.6$ \\
\bottomrule
\end{tabular}
\end{table}

For comparison with the matching predictions below, the four displayed best-fit coefficients are
\begin{equation}
  (C_W,C_{HW},C_{Hq}^{(3)},C_{tW})
  =
  (5.9 \pm 3.1,\ -3.0 \pm 8.3,\ 0.33 \pm 0.30,\ 1.5 \pm 0.9)~\tev^{-2}, 
\end{equation} 
with the corresponding profile intervals given in Table~\ref{tab:gstc_fullfit}. The coefficient $C_{Hq}^{(1)}$ is profiled in the fit with a best-fit value of $-0.27 \pm 0.29~\tev^{-2}$.

The residuals in Table~\ref{tab:bestfit_pred} show which discrepancies remain after profiling the five coefficients. The fitted $tWZ$ signal strength is $1.69$, compared with the measured $1.77\pm0.32$. The $tq\gamma$, $tH$ and $t\bar tH$ measurements retain pulls of approximately $+1.6\sigma$, $+1.7\sigma$ and $-1.8\sigma$, respectively. In this unbounded linear approach, the predicted signal strength for $tHq$ comes out very close to zero. 
The VBS pulls stay below half a standard deviation. The two triboson inputs share a single response, so the fit can only place its common prediction between them: it lands at $1.43$, leaving the ATLAS $VVZ$ pull at $+0.4$ and the CMS $WWZ$ pull at $-1.2$. For the five electroweak precision observables, the fit brings $A_{\rm FB}^{0,b}$ to its measured central value and leaves the remaining residuals small. 
We note that this result is specific to the five-observable subset. Independent measurements of the electron asymmetry $A_e$, which enters $A_{\rm FB}^{0,b} = \frac{3}{4} A_e A_b$, are not included in our fit.  
Since $A_b$ stays close to the SM value, the agreement in $A_{\rm FB}^{0,b}$ comes from a shift in $A_e$ to a value some $4\sigma$ below the LEP/SLD determination. A complete treatment of all $Z$-pole EWPOs would need an extended SMEFT setup with a larger number of operators, which we leave for future work. 

The heavy-fermion current sector developed in Section~\ref{sec:thresholds} does not generate $C_W$ at tree level. To test whether the fit depends critically on this purely bosonic direction, we therefore repeat it with $C_W$ fixed to zero. The other four coefficients remain free, and the best-fit values are
\begin{equation}
  (C_{HW},C_{Hq}^{(3)},C_{tW})
  =
  (6.6 \pm 6.6,\ 0.52 \pm 0.28,\ 2.0 \pm 0.8)~\tev^{-2},
\end{equation}
with $C_{Hq}^{(1)}=-0.46\pm0.27~\tev^{-2}$. This result is shown in red in Figure~\ref{fig:fit_result}. Fixing $C_W$ shifts $C_{Hq}^{(3)}$ and $C_{tW}$ by about $0.6\sigma$ of the full-fit uncertainties. The weakly determined $C_{HW}$ moves further, from $-3.0$ to $6.6~\tev^{-2}$, somewhat more than one standard deviation, but its interval remains broad and compatible with zero in both fits. The minimum is $\chi^2_{\min}=14.9$ for nine d.o.f., corresponding to $p=0.094$. Relative to the SM point, $\Delta\chi^2=9.8$. Counting the four free coefficients gives the formal Wilks conversion $p\simeq0.04$. The increase relative to the five-coefficient minimum is $\Delta\chi^2=3.7$, equivalent to about $1.9\sigma$ for one parameter.

The cost of imposing $C_W=0$ is shared between the two channels that constrain the triple-gauge direction. The $tWZ$ pull increases from $+0.2$ to $+1.4$, and the ATLAS $VVZ$ pull from $+0.4$ to $+1.4$: with $C_W$ free, the common triboson prediction sits at $1.43$, between the two triboson measurements, whereas fixing $C_W=0$ pulls it down to $1.05$, close to the CMS value. The electroweak precision observables change little, while the $tHq$ prediction moves away from zero to $\mu_{tHq} = 1.6$. Since the minimum $\chi^2$ increases by only $3.7$ units, the reduced fit remains compatible with the data and provides the starting point for the matching analysis.

\section{Gauge-invariant vector-like-quark interpretation}
\label{sec:thresholds}
The $C_W=0$ fit remains compatible with the data, while $C_{HW}$ is only weakly determined. We therefore isolate the current and dipole directions relevant to a heavy-fermion interpretation and ask which of them can be generated by the sector below.

The current and dipole impose different requirements on this sector. Mixing $Q_3$ with a charge-$2/3$ singlet generates $C_{Hq}^{(1,3)}$ at tree level. The opposite chirality needed for the dipole is supplied by coupling $t_R$ to a hypercharge-$1/6$ doublet. Since the singlet and doublet are inequivalent representations, a single vector-like multiplet cannot provide both couplings within this ansatz and in the absence of direct dimension-three heavy--light mixing. Moreover, the singlet contribution to $C_{Hq}^{(3)}$ has the wrong sign for the positive fit direction. We therefore add triplet partners that can reverse this sign while preserving the $Zb_L\bar b_L$-safe combination. The dipole requires additional loop dynamics, which we discuss below~\cite{Kaplan:1991dc,DeSimone:2012fs,Panico:2015jxa}. The current sector generates neither $C_W$ nor $C_{HW}$ at tree level, so an independent sizeable bosonic contribution would require further dynamics.

\subsection{Heavy-fermion currents and custodial alignment}
\label{sec:gauge-current-sector}
These requirements lead to the field content used in the current matching: a singlet $U$ for the $Q_3$ mixing, a doublet ${\cal Q}$ for the $t_R$ coupling, and two triplets whose contributions can be aligned in the $Zb_L\bar b_L$-safe direction:
\begin{equation}
 \begin{aligned}
  U&\sim(\mathbf 3,\mathbf 1)_{2/3}, &
  {\cal Q}&\sim(\mathbf 3,\mathbf 2)_{1/6},\\
  \Sigma_1&\sim(\mathbf 3,\mathbf 3)_{-1/3}, &
  \Sigma_2&\sim(\mathbf 3,\mathbf 3)_{2/3}.
 \end{aligned}
  \label{eq:heavy_reps}
\end{equation}
All fermions are vector-like, and the subscripts denote hypercharge. The doublet is ${\cal Q}=(T,B)$; $\Sigma_2$ contains a charge-$5/3$ state and $\Sigma_1$ a charge-$-4/3$ state. These exotic partners are part of the spectrum tested by direct searches. The fields shown in Eq.~\eqref{eq:heavy_reps} are those used in the matching, not a complete custodial representation. A custodial completion of the embedding considered here generally contains further partners, including a $(\mathbf 3,\mathbf 2)_{7/6}$ doublet~\cite{DeSimone:2012fs,Aguilar-Saavedra:2013qpa}.

To isolate the tree-level origin of the current operators, we retain the heavy--light Yukawa interactions and work in a basis without direct dimension-three heavy--light mass mixing:
\begin{equation}
\begin{split}
{\cal L}_{\rm cur}={}&
 \bar U(i\slashed D-M_U)U
 +\bar{\cal Q}(i\slashed D-M_{\cal Q}){\cal Q}
 +\sum_{r=1,2}\bar\Sigma_r(i\slashed D-M_{\Sigma_r})\Sigma_r\\
&-\bigg[
 y_L\,\bar Q_3\widetilde H U_R
 +y_R\,\bar{\cal Q}_L\widetilde H t_R
 +\frac{\lambda_1}{2}\bar Q_3\tau^aH\Sigma_{1R}^{a}
 +\frac{\lambda_2}{2}\bar Q_3\tau^a\widetilde H\Sigma_{2R}^{a}
 +{\rm h.c.}\bigg].
\end{split}
\label{eq:current_lagrangian}
\end{equation}
The $y_L$ term mixes $Q_3$ with the singlet and generates the left-handed current operators. The $y_R$ term couples $t_R$ to the doublet, supplies the second chirality needed for the dipole, and generates $C_{Hu}$. Because $U$ and ${\cal Q}$ belong to different electroweak representations, no vector-like mass term connects them; an explicit dipole completion must introduce an additional heavy--heavy interaction or another mediator. The $\lambda_{1,2}$ terms provide the triplet contributions needed below. For the numerical benchmark, we set $M_U=M_{\cal Q}\equiv M_F$ to reduce the number of independent mass ratios; this degeneracy is not required by gauge symmetry.

The singlet lies on the $Zb_L\bar b_L$-safe direction, $C_{Hq}^{(1)}+C_{Hq}^{(3)}=0$, but contributes negatively to $C_{Hq}^{(3)}$~\cite{Agashe:2006at}. The fit instead points towards a positive value. The triplets can reverse the sign, but a generic triplet contribution would move the theory away from the protected combination. We therefore include both triplets and correlate their masses and couplings so that their net contribution raises $C_{Hq}^{(3)}$ while maintaining $C_{Hq}^{(1)}+C_{Hq}^{(3)}\simeq0$ at the matching scale. Such a relation can arise from approximate custodial symmetry, although the active fields in Eq.~\eqref{eq:heavy_reps} do not by themselves form a complete $SU(2)_L\times SU(2)_R$ representation~\cite{Contino:2008hi,Mrazek:2009yu,Matsedonskyi:2012ym,Azatov:2013hya,Xie:2019gya}.

Integrating out the heavy fermions makes the sign and alignment conditions explicit. We use $\order_{Hu}=(H^\dagger i\overleftrightarrow D_\mu H)(\bar t_R\gamma^\mu t_R)$. At tree level and to leading order in $1/M^2$, one obtains~\cite{deBlas:2017xtg,Carmona:2021xtq,Crivellin:2022fdf,Guedes:2024vuf}
\begin{equation}
\label{eq:match}
\begin{split}
  C_{Hq}^{(3)}
  &\simeq -\frac{|y_L|^2}{4M_U^2}
  +\frac{|\lambda_1|^2}{16M_{\Sigma_1}^2}
  +\frac{|\lambda_2|^2}{16M_{\Sigma_2}^2},\\
  C_{Hq}^{(1)}
  &\simeq +\frac{|y_L|^2}{4M_U^2}
  -\frac{3|\lambda_1|^2}{16M_{\Sigma_1}^2}
  +\frac{3|\lambda_2|^2}{16M_{\Sigma_2}^2},\\
  C_{Hu}
  &\simeq -\frac{|y_R|^2}{2M_{\cal Q}^2}.
\end{split}
\end{equation}
The coefficient $C_{Hu}$ is not varied in the fit, but it is retained here because the same right-handed mixing modifies the $Zt_R\bar t_R$ coupling and enters the correlated constraints. Adding the first two matching relations isolates the combination constrained by $Zb_L\bar b_L$:
\begin{equation}
 C_{Hq}^{(1)}+C_{Hq}^{(3)}
 \simeq -\frac{|\lambda_1|^2}{8M_{\Sigma_1}^2}
 +\frac{|\lambda_2|^2}{4M_{\Sigma_2}^2}.
 \label{eq:custodial_alignment}
\end{equation}
At the matching scale, the $Zb_L\bar b_L$-safe direction therefore fixes the relative size of the two triplet contributions.

\subsection{Dipole and bosonic matching}
A dipole connects the external fields $Q_3$ and $t_R$. In Eq.~\eqref{eq:current_lagrangian} they couple to different heavy fermions, $U$ and ${\cal Q}$, so the loop cannot be closed without an interaction connecting the two multiplets or an additional bosonic mediator. In a weakly coupled completion, the result has the form
\begin{equation}
 C_{tW}(\mu_M)=\frac{g_2}{16\pi^2M_*^2}\,
 {\cal F}_{tW}(\text{heavy couplings and mass ratios}),
 \label{eq:dipole_general}
\end{equation}
where $\mu_M$ is the matching scale, $M_*$ denotes the characteristic mass in the loop, and the form factor ${\cal F}_{tW}$ is fixed once the additional interactions and heavy mass ratios are specified. In a specified completion, the same one-loop matching would also determine $C_{tB}$, $C_{tG}$ and correlated Higgs and flavour coefficients; $C_{tW}$ is therefore not an independent parameter in a complete model. Since we do not choose a particular dipole completion here, we parametrise its net contribution in the numerical scan as
\begin{equation}
 C_{tW}^{\rm bench}\equiv
 \kappa_{\rm dip}^{\rm eff}
 \frac{g_*y_Ly_R}{16\pi^2M_F^2}\frac{M_F}{m_t}.
 \label{eq:kappadip_definition}
\end{equation}
In this benchmark parametrisation, $g_*$ denotes a representative coupling connecting the heavy states, while $\kappa_{\rm dip}^{\rm eff}$ absorbs $g_2$ from Eq.~\eqref{eq:dipole_general}, the electroweak group factors, the loop function and the remaining mass-ratio dependence. The explicit factor $M_F/m_t$ is a broken-phase parametrisation of a possible heavy-line chirality enhancement; only the full product in Eq.~\eqref{eq:kappadip_definition} has physical meaning. The current Lagrangian leaves this quantity free, and the fit-preferred $C_{tW}$ is substantially larger than the reference value used below. Reproducing the fit-preferred dipole is therefore the main challenge for an explicit weakly coupled completion.

The reference benchmark focuses on the fermionic current and dipole sectors. An independent sizeable $C_{HW}$ is not generated at tree level by this current sector; to illustrate how an additional source would enter, we introduce the following low-energy scalar parametrisation:
\begin{equation}
 {\cal L}_S=\frac12(\partial_\mu S)^2-\frac12M_S^2S^2
 +\kappa_HM_S S H^\dagger H
 +\frac{\kappa_{WW}}{M_S}S W^a_{\mu\nu}W^{a\mu\nu}.
 \label{eq:scalar_threshold}
\end{equation}
Integrating out $S$ shows explicitly that a separate source of $C_{HW}$ also affects the Higgs sector. At tree level it gives $C_{HW}=\kappa_H\kappa_{WW}/M_S^2$ and, with $\order_{H\Box}=(H^\dagger H)\Box(H^\dagger H)$, $C_{H\Box}=-\kappa_H^2/(2M_S^2)$, together with the renormalisable quartic shift $+\kappa_H^2(H^\dagger H)^2/2$. In this scalar parametrisation, $C_{HW}$ is accompanied by $C_{H\Box}$ and a shift of the Higgs quartic coupling. Moreover, the $S W_{\mu\nu}W^{\mu\nu}$ interaction is dimension five and must itself arise from an additional charged threshold. We omit this sector from the reference point; its purpose here is only to show how a sizeable $C_{HW}$ would enter, while the fermion current sector gives $C_W=0$ at tree level.

\subsection{Reference point and correlated constraints}
The reference point should lie where the current operators remain appreciable and direct searches are already relevant. We therefore choose masses near present search sensitivity and ask whether the resulting matching can reproduce the positive best-fit current direction while preserving the $Zb_L\bar b_L$-safe combination. The point is not intended to reproduce the full SMEFT best fit. The reference masses are
\begin{equation}
M_U\simeq M_{\cal Q}\equiv M_F\simeq 1.60~\tev,
\qquad
M_{\Sigma_1}\simeq 2.0~\tev,
\qquad
M_{\Sigma_2}\simeq 1.9~\tev,
\end{equation}
with effective dipole parameters
\begin{equation}
\kappa_{\rm dip}^{\rm eff}\simeq 2,\qquad g_*\simeq 1.3,
\end{equation}
and fermion couplings
\begin{equation}
y_L \simeq 1.2, \quad y_R \simeq 1.4,\quad
\lambda_{1} \simeq 3.3,\quad \lambda_{2}\simeq 2.2.
\end{equation}
The couplings $y_L$, $y_R$ and $\lambda_{1,2}$ specify the tree-level current sector. They are chosen to give a positive $C_{Hq}^{(3)}$ while nearly cancelling $C_{Hq}^{(1)}+C_{Hq}^{(3)}$. By contrast, $g_*\kappa_{\rm dip}^{\rm eff}$ specifies the strength required from the unresolved dipole completion rather than a prediction of the current Lagrangian.

The tree-level coefficients confirm the intended alignment. Substituting the masses and couplings into Eq.~\eqref{eq:match} gives
\begin{equation}
 (C_{Hq}^{(3)},C_{Hq}^{(1)},C_{Hu})_{\rm tree}
 \simeq(0.113,-0.118,-0.383)~\tev^{-2},
 \qquad
 C_{Hq}^{(1)}+C_{Hq}^{(3)}\simeq-0.0051~\tev^{-2}.
 \label{eq:benchmark_tree_values}
\end{equation}
For the numerical comparisons, we use $(0.100,-0.100,-0.385)~\tev^{-2}$ for these three entries, as in the mass-basis calculation. These rounded assignments are close to the leading values in Eq.~\eqref{eq:benchmark_tree_values}; their difference is not interpreted as a higher-order matching correction.

For comparison with the fit, the dominant Wilson coefficients at the reference point are
\begin{equation}
(C_W,C_{HW},C_{Hq}^{(3)},C_{tW}) \simeq
\left(0,0,0.1,0.1\right)~\tev^{-2}.
\end{equation}
For the numerical comparison, we use the fixed-input convention of Section~\ref{sec:response-map}. Equations~\eqref{eq:match} and~\eqref{eq:custodial_alignment} are matching-scale relations, but no RG map between the heavy thresholds and the collider coefficients is applied. The reference $C_{tW}\simeq0.1~\tev^{-2}$ lies inside the nominal $95\%$ interval in Table~\ref{tab:gstc_fullfit}, near its lower end, and well below the best-fit value. The point is therefore not a best-fit UV reconstruction; it tests whether a spectrum at $1.6$--$2.0~\tev$ can reproduce the fitted signs and current alignment while making the remaining dipole requirement explicit.

The associated mass-basis benchmark also modifies the top Yukawa coupling, making $t\bar tH$ an immediate correlated test. For this point,
\begin{equation}
\kappa_t \simeq 0.91,
\end{equation}
which specifies the Higgs coupling beyond Eq.~\eqref{eq:current_lagrangian}. In the $\kappa$ framework~\cite{LHCHiggsCrossSectionWorkingGroup:2013rie}, this corresponds to $\mu_{t\bar tH}\simeq\kappa_t^2\simeq0.83$ when branching fractions and acceptances are held fixed. The shift follows the below-SM central values in Table~\ref{tab:measurements} and in the recent CMS Higgs combination~\cite{CMS:2026nce}; it is an associated benchmark value and is not added separately to the likelihood.

\begin{table}[t]
\centering
\caption{Wilson coefficients used for the reference point. The current entries reproduce the signs and alignment of the gauge-invariant tree sector in Eq.~\eqref{eq:current_lagrangian}; their leading heavy-mass values are given in Eq.~\eqref{eq:benchmark_tree_values}. The remaining loop-level entries are fixed benchmark values. A complete one-loop matching calculation, including $C_{tB}$ and $C_{tG}$, requires a specified dipole completion. Scalar contributions are omitted.}
\label{tab:reference-wilsons}
\begin{tabular}{l r l r}
    \toprule
    Coefficient & Value [$\tev^{-2}$]
      & Coefficient & Value [$\tev^{-2}$] \\
    \midrule
    $C_{Hq}^{(3)}$ & $+1.00\times10^{-1}$
      & $C_{Hq}^{(1)}$ & $-1.00\times10^{-1}$ \\
    $C_{Hu}$ & $-3.85\times10^{-1}$
      & $C_{tW}$ & $+9.97\times10^{-2}$ \\
    $C_{G}$ & $-8.61\times10^{-5}$
      & $C_{HG}$ & $+4.29\times10^{-3}$ \\
    $C_{HW}$ & $+2.34\times10^{-3}$
      & $C_{HB}$ & $+3.38\times10^{-5}$ \\
    $C_{HWB}$ & $+4.09\times10^{-5}$
      & $C_{W}$ & $-5.99\times10^{-5}$ \\
    \bottomrule
  \end{tabular}
\end{table}

Gauge invariance also makes the benchmark testable through correlated $t\bar tZ$ and $Wtb$ couplings. For the same mass-basis point, the neutral-current shifts are
\begin{equation}
\delta_R(t\bar tZ) = -7.7\%,\quad \delta_L(t\bar t Z)=+1.6\%,
\end{equation}
while the charged-current shifts are
\begin{equation}
\bar\delta_L(t\bar bW^-) =+0.56\%,\quad \bar\delta_R(t\bar bW^-) = \frac{g_{t\bar bW^-}^R}{\left[g_{t\bar bW^-}^L\right]_{\text{SM}}} \simeq 10^{-5}.
\end{equation}
These shifts cannot be assessed by comparing them one by one with inclusive rate precisions. Current $t\bar tZ$ measurements have a typical precision of $5$--$7\%$~\cite{CMS:2025tre,CMS:2024mke,ATLAS:2023eld}, but the opposite chiral shifts partly cancel in the inclusive rate. The ATLAS full-Run~2 result $\sigma(tq+\bar tq)=221\pm13~{\rm pb}$~\cite{ATLAS:2024ojr} corresponds to an approximately $3\%$ sensitivity to a universal rescaling of the left-handed $Wtb$ coupling when the top quark width, branching fractions and other EFT contributions are fixed. The measured right-handed helicity fraction $f_R=-0.002\pm0.014$~\cite{ATLAS:2022rms} depends nonlinearly on the vector and tensor $Wtb$ couplings and is not itself a measurement of $\bar\delta_R$. A joint constraint on the two chiral shifts therefore requires the relevant differential information.

\begin{figure}[!t]
\centering
\includegraphics[width=0.7\textwidth]{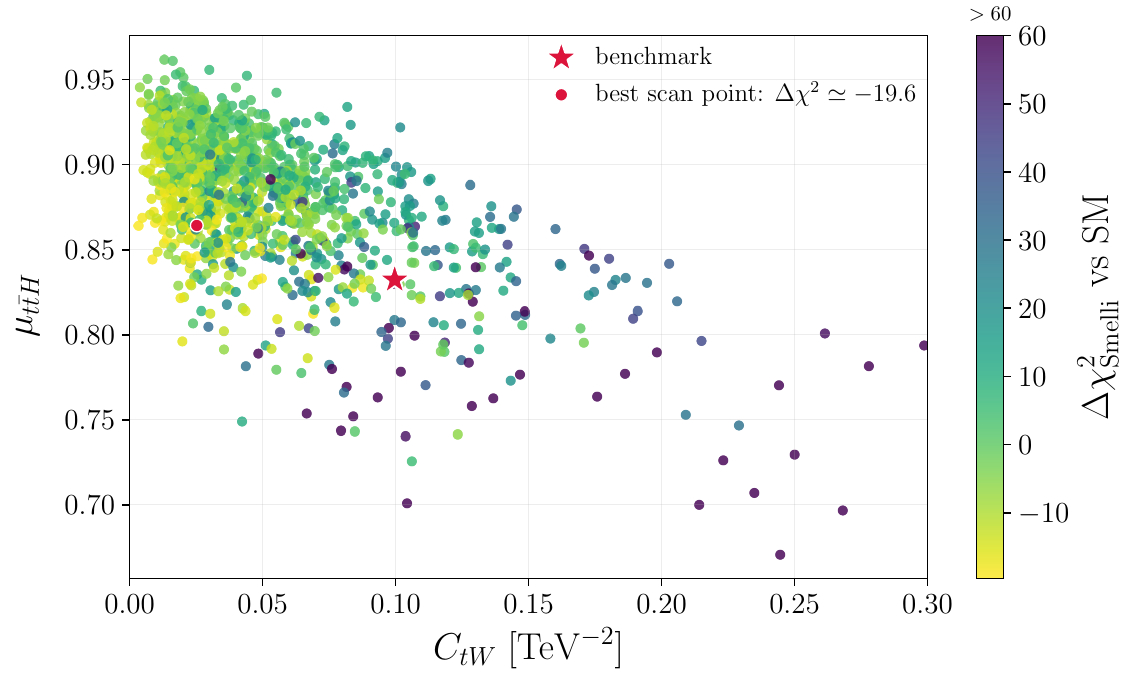}
\caption{\label{fig:scatter} Correlation of $\mu_{t\bar tH}$ and $C_{tW}$ in the benchmark scan, coloured by $\Delta\chi^2_{\rm smelli}=\chi^2_{\rm point}-\chi^2_{\rm SM}$. Negative values denote an improvement over the SM. The annotated $-18.6$ marks the best scanned point, not the reference point discussed in the text.}
\end{figure}

A single reference point does not show whether the alignment survives reasonable variations of the masses and couplings. We therefore vary the heavy masses independently in $[1.6,3]~\tev$, with $\kappa_{\rm dip}^{\rm eff}\in[1,6]$, $y_L\in[0.40,3.50]$, $y_R\in[0.45,3.5]$ and $\lambda_{1,2}\in[-3.5,3.5]$. The resulting sample covers $C_{Hq}^{(1)}+C_{Hq}^{(3)}\in[-0.25,0.25]~\tev^{-2}$ around the aligned direction and exposes the correlation between the effective dipole, the top Yukawa and the selected flavour likelihood shown in Figure~\ref{fig:scatter}. The flavour quantity used for the colour scale is the selected-block EFT check defined below. These points are evaluated with Eq.~\eqref{eq:kappadip_definition} and the mass-basis map employed in the numerical calculation. Since no explicit one-loop matching condition is imposed, they test the low-energy compatibility of the parametrisation rather than establishing a complete dipole model.

The same masses and mixings are probed directly through production of the vector-like partners. Existing limits already probe the reference mass range, although their translation is benchmark dependent. Searches by the ATLAS Collaboration exclude $m_T<1.70~\tev$ for $\mathrm{BR}(T\to Wb)=1$ and $m_T<1.36~\tev$ for the singlet-like pattern $Wb:Ht:Zt=1/2:1/4:1/4$~\cite{ATLAS:2024vlqpair}. The $1.56~\tev$ limit in the $Ht$ analysis~\cite{ATLAS:2026ojf} and the $1.46~\tev$ limit in the $Zt$ analysis~\cite{ATLAS:2022hnn} apply to their stated doublet benchmarks. The CMS Collaboration excludes masses below at least $1.48~\tev$ over the third-generation branching-fraction plane, with a reach up to $1.54~\tev$ in the most sensitive regions~\cite{CMS:2022fck}.

Applying these limits to the reference point requires its branching fractions and a translation between coupling conventions. For the $T$-like state used in the mass-basis study we take $Wb:Ht:Zt\simeq52:17:31$. This assignment specifies that state beyond the truncated current Lagrangian, which does not contain the complete charge-$2/3$ mass matrix. Its reference mass of $1.6~\tev$ lies close to current pair-production sensitivity. Single-production searches provide a complementary constraint: the quoted ATLAS $Ht/Zt$ interpretation reaches $\kappa\simeq0.35$ near $1.6~\tev$ and excludes masses below about $2.3~\tev$ for $\kappa\gtrsim0.53$~\cite{ATLAS:2023singleT}. Dedicated $Wb$ searches are directly relevant. For its singlet-$T$ benchmark, an ATLAS analysis obtains limits on $\kappa$ between $0.22$ and $0.52$ over masses from $1.15$ to $2.3~\tev$~\cite{ATLAS:2025singleWb}. For an exclusive $Wb$ decay, the CMS Collaboration reaches $\kappa_W=0.086$ near $1.4~\tev$ and, for $\kappa_W=0.2$, excludes masses below $2.4~\tev$~\cite{CMS:2026singleWb}. The mixing estimate
\begin{equation}
 \kappa_{\rm eff}\equiv\frac{y_Lv}{\sqrt{2}M_F}\simeq0.13
\end{equation}
sets the expected scale, although its relation to either experimental coupling convention depends on the representation, chirality, branching fractions and width. A multi-representation sector contains several nearby states, including exotic electric charges, and must be tested as a complete spectrum. If $T\to tS$ is open, it dilutes the conventional branching fractions while producing a dedicated exotic-decay signature. These searches place the reference point squarely in the region now being tested. A definite exclusion statement requires a recast with the complete masses, widths, branching fractions and coupling dictionary.

The flavour sector probes a different assumption: sizeable heavy--light mixing must be aligned with the third generation if it is not to induce strongly constrained light-generation transitions. We assume either approximate minimal flavour violation or a $U(2)^3$ structure~\cite{DAmbrosio:2002vsn,Barbieri:2012uh}, with negligible direct couplings to the first two generations. More general couplings feed $B_{s,d}$ mixing, $b\to s\gamma$, $b\to s\ell^+\ell^-$ and $b\to s\nu\bar\nu$, together with kaon and charm mixing once light-quark interactions are present.

We do not want to overreach with the flavour conclusion for this benchmark. An affirmative statement would require a complete specification of the flavour structure and dipole-generating sector, matching at a stated scale, renormalisation-group evolution and a clearly defined flavour likelihood, including finite one-loop effects.

The construction can reproduce the fit-preferred current-operator direction, but it does not generate the dipole at tree level. In particular, tree-level matching can yield a positive $C_{Hq}^{(3)}$ while preserving the heavy-scale alignment $C_{Hq}^{(1)}+C_{Hq}^{(3)}\simeq0$; the preferred dipole must arise from additional dynamics. Figure~\ref{fig:scatter} shows how, within the low-energy scan, the effective dipole and the suppression of $\mu_{t\bar tH}$ vary away from the aligned point, with the colour scale indicating the selected flavour likelihood. Because the dipole has not been matched explicitly at one loop, this pattern is a consistency check rather than a UV prediction. Improved $t\bar tH$ and electroweak single-top likelihoods can probe these effects indirectly, while searches for the singlet, doublet, triplet and exotic-charge partners test the threshold spectrum directly. Together, these measurements will test whether a complete model of the required dipole dynamics remains viable.

\section{Summary and outlook}
The measurements considered here show an intriguing, if still developing, pattern. The central values in $tWZ$ and $tq\gamma$ lie above their Standard Model predictions, and the multilepton analysis admits a large $tHq$ contribution together with a $t\bar tH$ rate below the SM expectation. At the same time, the VBS and triboson channels that probe related electroweak interactions remain close to the SM. These observations should not be viewed as unrelated numbers. They involve the same top quark, Higgs and gauge currents, and gauge invariance forces any common explanation to appear in several channels at once. The question posed in this paper is whether these modest shifts can be the first signs of such a common deformation.

The five-parameter SMEFT fit gives an affirmative answer at the level of the selected likelihood. The main effects are a positive pure triple-gauge coefficient $C_W$ and a weak top dipole coefficient, $C_{tW}$, which have a pull from zero at about $2\sigma$. These are accompanied by a positive best-fit value of the left-handed current coefficient $C_{Hq}^{(3)}$. The coefficient $C_{Hq}^{(1)}$ is needed alongside it because precision $Zb_L\bar b_L$ data constrain the sum $C_{Hq}^{(1)}+C_{Hq}^{(3)}$. The coefficient $C_{HW}$ is poorly determined. Quantitatively, we find
\begin{equation}
 \chi^2_{\min,\rm SM}=24.6\quad(13~{\rm d.o.f.}),\qquad
 \chi^2_{\min,\rm SMEFT}=11.2\quad(8~{\rm d.o.f.}),
\end{equation}
or $\Delta\chi^2=13.4$ for five additional coefficients. The corresponding local likelihood-ratio $p$-value is $0.02$. Setting $C_W=0$ still gives $\chi^2_{\min}=14.9$ for nine degrees of freedom, $3.7$ above the full minimum. Thus the improvement does not rely on a large pure triple-gauge interaction and can be discussed in terms of the current and dipole sectors. Within the five $Z$-pole observables included here, the same fit also removes the residual in $A_{\rm FB}^{0,b}$ without producing large pulls in the other four observables. This last statement is deliberately limited to that subset: the independent leptonic asymmetries and the complete set of EWPOs must be included before drawing a conclusion about the global electroweak fit. 

The EFT fit is only the first half of the problem. Its coefficients must be generated by heavy fields that respect the same gauge symmetry and that survive electroweak precision, Higgs, flavour and direct-search constraints. At leading order in the weakly coupled renormalisable matching considered here, a one-particle simplified model does not provide the required ingredients. A scalar field can generate bosonic operators but not the protected third-generation current structure at tree level. A vector can generate currents at tree level but not the weak dipole, and a consistent gauge completion brings further states. A single vector-like-quark multiplet also fails within our heavy--light Yukawa ansatz: the coupling to $Q_3$ requires a charge-$2/3$ singlet, whereas the coupling to $t_R$ requires a hypercharge-$1/6$ doublet. The positive current direction then calls for additional custodial partners. The explanation is therefore intrinsically multi-field and, unless the masses are degenerate, multi-threshold.

The viable construction that emerges from this exercise contains a singlet $U$, a doublet ${\cal Q}$ and two triplets whose contributions are aligned so that
\begin{equation}
 C_{Hq}^{(1)}+C_{Hq}^{(3)}\simeq0
\end{equation}
at the matching scale. This approximate custodial relation protects $Zb_L\bar b_L$ while allowing a positive $C_{Hq}^{(3)}$. The current operators are generated at tree level. The weak dipole has a different origin: the singlet and doublet sectors must be connected through additional heavy--heavy interactions or a further mediator, and the matching is loop induced. Once this dynamics is specified, it will also generate correlated dipole, Higgs and flavour operators. A sizeable independent $C_{HW}$ would similarly point to an additional charged or bosonic sector.

Our aligned reference point shows that these requirements can be realised in a spectrum close to present LHC sensitivity. For partner masses between $1.6$ and $2.0~\tev$, it gives
\begin{equation}
 (C_W,C_{HW},C_{Hq}^{(3)},C_{tW})
 \simeq(0,0,0.1,0.1)~\tev^{-2},
\end{equation}
with the required sign and $Zb_L\bar b_L$ alignment in the current sector. The associated mass-basis point has $\kappa_t\simeq0.91$ and percent-level shifts in the weak top currents, while the aligned-flavour check gives the indicative value $\Delta\chi^2_{\rm flavour}\simeq0.6$. This benchmark is not the SMEFT fit minimum: in particular, its dipole is well below the best-fit value. It shows, at the level of the checks performed here, that the gauge-current part of the construction can satisfy the correlated constraints without being decoupled from experiment. The remaining task is to specify the loop dynamics that produces the larger dipole. Current vector-like-quark searches already probe this mass range, but a firm statement requires the complete spectrum, widths, branching fractions and coupling conventions, including the charge-$5/3$ and charge-$-4/3$ partners.

An important lesson is that this interpretation cannot be tested one constraint at a time. Partners introduced to protect $Zb_L\bar b_L$ also modify $Wtb$, $t\bar tZ$ and Higgs couplings. Interactions added to generate the dipole feed further collider and flavour operators. Raising the masses to avoid a direct-search limit suppresses the Wilson coefficients and must be compensated elsewhere. If the present deviations become more significant, the fit, the matching and the indirect and direct constraints will have to be revisited together as the allowed spectra are narrowed down.

The most useful next step is therefore a likelihood-level experimental test of these correlations. Public likelihoods or covariance information for the joint $tHq$--$t\bar tH$ extraction, the VBS combination and the triboson measurements, together with a $VH$-subtracted breakdown of $VVZ$ into its $WWZ$, $WZZ$ and $ZZZ$ components, fiducial or differential response information for $tq\gamma$ and energy-resolved information for $tWZ$, would remove the leading approximations in the present analysis. More precise $t\bar tH$, $t\bar tZ$, $W$-helicity and electroweak single top-quark measurements would test the correlated couplings directly. On the search side, the appropriate target is not an isolated $T$ benchmark. Dedicated pair- and single-production interpretations should cover singlet, doublet and triplet partners, mixed $Wb$, $Zt$ and $Ht$ branching fractions, several nearby mass eigenstates and the exotic charge $5/3$ and charge $-4/3$ states. If an additional boson participates in the dipole completion, searches for cascades such as $T\to tS$ provide a complementary handle. Should the present rate shifts persist, the new-physics target will be a correlated current--dipole deformation accompanied by a structured partner spectrum. We encourage the experimental collaborations to test the two together.

\section*{Acknowledgements}
%
We thank Juliette Alimena for helpful comments on the manuscript.
C.E. thanks the Institute for Theoretical Physics at KIT for hospitality during various stages of this work.
R.K. acknowledges support from the Deutsche Forschungsgemeinschaft (DFG, German Research Foundation) under Germany's Excellence Strategy -- EXC 2121 ``Quantum Universe'' -- 390833306.
%

\bibliographystyle{JHEP_mod}
\bibliography{references}

\providecommand{\href}[2]{#2}\begingroup\raggedright\begin{thebibliography}{10}

\bibitem{CMS:2025tre}
CMS Collaboration, \emph{{Observation of tWZ production at the CMS
  experiment}}, \href{https://doi.org/10.1103/rk6w-1pcl}{\emph{Phys. Rev.
  Lett.} 136 (2026) 081802},
  [\href{https://arxiv.org/abs/2510.19080}{{\ttfamily 2510.19080}}].

\bibitem{ATLAS:2023qdu}
ATLAS Collaboration, \emph{{Observation of single-top-quark production in
  association with a photon using the ATLAS detector}},
  \href{https://doi.org/10.1103/PhysRevLett.131.181901}{\emph{Phys. Rev. Lett.}
  131 (2023) 181901}, [\href{https://arxiv.org/abs/2302.01283}{{\ttfamily
  2302.01283}}].

\bibitem{ATLAS:2025eua}
ATLAS Collaboration, \emph{{Measurement of the Higgs boson production in
  association with top quarks in multilepton final states in pp collisions at $
  \sqrt{s}=13 $ TeV with the ATLAS detector}},
  \href{https://doi.org/10.1007/JHEP05(2026)183}{\emph{JHEP} 05 (2026) 183},
  [\href{https://arxiv.org/abs/2510.23755}{{\ttfamily 2510.23755}}].

\bibitem{CMS:2026add}
CMS Collaboration, \emph{{Combination of vector boson scattering measurements
  in leptonic final states in proton-proton collisions at $\sqrt{s}$ = 13
  TeV}},  \href{https://arxiv.org/abs/2608.01289}{{\ttfamily 2608.01289}}.

\bibitem{ATLAS:2024nab}
ATLAS Collaboration, \emph{{Observation of $VVZ$ production at $\sqrt{s}=13$
  TeV with the ATLAS detector}},
  \href{https://doi.org/10.1016/j.physletb.2025.139527}{\emph{Phys. Lett. B}
  866 (2025) 139527}, [\href{https://arxiv.org/abs/2412.15123}{{\ttfamily
  2412.15123}}].

\bibitem{CMS:2025hlu}
CMS Collaboration, \emph{{Measurement of WWZ and ZH production cross sections
  at $\sqrt{s}=13$ and 13.6~TeV}},
  \href{https://doi.org/10.1103/6z3d-zjw4}{\emph{Phys. Rev. Lett.} 135 (2025)
  091802}, [\href{https://arxiv.org/abs/2505.20483}{{\ttfamily 2505.20483}}].

\bibitem{Grzadkowski:2010es}
B.~Grzadkowski, M.~Iskrzynski, M.~Misiak and J.~Rosiek, \emph{{Dimension-six
  terms in the Standard Model Lagrangian}},
  \href{https://doi.org/10.1007/JHEP10(2010)085}{\emph{JHEP} 10 (2010) 085},
  [\href{https://arxiv.org/abs/1008.4884}{{\ttfamily 1008.4884}}].

\bibitem{Brivio:2020onw}
I.~Brivio, \emph{{SMEFTsim 3.0 --- a practical guide}},
  \href{https://doi.org/10.1007/JHEP04(2021)073}{\emph{JHEP} 04 (2021) 073},
  [\href{https://arxiv.org/abs/2012.11343}{{\ttfamily 2012.11343}}].

\bibitem{Ellis:2018gqa}
J.~Ellis, C.~W. Murphy, V.~Sanz and T.~You, \emph{{Updated global SMEFT fit to
  Higgs, diboson and electroweak data}},
  \href{https://doi.org/10.1007/JHEP06(2018)146}{\emph{JHEP} 06 (2018) 146},
  [\href{https://arxiv.org/abs/1803.03252}{{\ttfamily 1803.03252}}].

\bibitem{Biekotter:2018ohn}
A.~Biek{\"o}tter, T.~Corbett and T.~Plehn, \emph{{The gauge-Higgs legacy of the
  LHC Run II}},
  \href{https://doi.org/10.21468/SciPostPhys.6.6.064}{\emph{SciPost Phys.} 6
  (2019) 064}, [\href{https://arxiv.org/abs/1812.07587}{{\ttfamily
  1812.07587}}].

\bibitem{Madigan:2022cvc}
M.~Madigan, \emph{{Top, Higgs, diboson and electroweak fit to the Standard
  Model effective field theory}},  in \emph{{15Th international workshop on top
  quark physics}}, 12, 2022, \href{https://arxiv.org/abs/2212.00384}{{\ttfamily
  2212.00384}}.

\bibitem{ALEPH:2005ema}
{ALEPH, DELPHI, L3, OPAL, SLD Collaborations, the LEP Electroweak Working
  Group, the SLD Electroweak and Heavy Flavour Working Groups},
  \emph{{Precision electroweak measurements on the Z resonance}}, {\emph{Phys.
  Rept.} 427 (2006) 257},
  [\href{https://arxiv.org/abs/hep-ex/0509008}{{\ttfamily hep-ex/0509008}}].

\bibitem{Janot:2019oyi}
P.~Janot and S.~Jadach, \emph{{Improved Bhabha cross section at LEP and the
  number of light neutrino species}},
  \href{https://doi.org/10.1016/j.physletb.2020.135319}{\emph{Phys.\ Lett.\ B}
  803 (2020) 135319}, [\href{https://arxiv.org/abs/1912.02067}{{\ttfamily
  1912.02067}}].

\bibitem{Bernreuther:2016ccf}
W.~Bernreuther, L.~Chen, O.~Dekkers, T.~Gehrmann and D.~Heisler, \emph{{The
  forward-backward asymmetry for massive bottom quarks at the $Z$ peak at
  next-to-next-to-leading order QCD}},
  \href{https://doi.org/10.1007/JHEP01(2017)053}{\emph{JHEP} 01 (2017) 053},
  [\href{https://arxiv.org/abs/1611.07942}{{\ttfamily 1611.07942}}].

\bibitem{Catani:1999nf}
S.~Catani and M.~H. Seymour, \emph{{Corrections of $\Order(\alpha_s^2)$ to the
  forward-backward asymmetry}},
  \href{https://doi.org/10.1088/1126-6708/1999/07/023}{\emph{JHEP} 07 (1999)
  023}, [\href{https://arxiv.org/abs/hep-ph/9905424}{{\ttfamily
  hep-ph/9905424}}].

\bibitem{Djouadi:1989uk}
A.~Djouadi, J.~H. Kuhn and P.~Zerwas, \emph{{B jet asymmetries in $Z$ decays}},
  \href{https://doi.org/10.1007/BF01621029}{\emph{Z.\ Phys.\ C} 46 (1990) 411}.

\bibitem{Abe:2000dq}
SLD Collaboration, \emph{{A high precision measurement of the left-right Z
  boson cross-section asymmetry}},
  \href{https://doi.org/10.1103/PhysRevLett.84.5945}{\emph{Phys.\ Rev.\ Lett.}
  84 (2000) 5945}, [\href{https://arxiv.org/abs/hep-ex/0004026}{{\ttfamily
  hep-ex/0004026}}].

\bibitem{Abe:2000uc}
SLD Collaboration, \emph{{First direct measurement of the parity violating
  coupling of the $Z^0$ to the $s$ quark}},
  \href{https://doi.org/10.1103/PhysRevLett.85.5059}{\emph{Phys.\ Rev.\ Lett.}
  85 (2000) 5059}, [\href{https://arxiv.org/abs/hep-ex/0006019}{{\ttfamily
  hep-ex/0006019}}].

\bibitem{Abe:2000hk}
SLD Collaboration, \emph{{An improved direct measurement of leptonic coupling
  asymmetries with polarized $Z$ bosons}},
  \href{https://doi.org/10.1103/PhysRevLett.86.1162}{\emph{Phys.\ Rev.\ Lett.}
  86 (2001) 1162}, [\href{https://arxiv.org/abs/hep-ex/0010015}{{\ttfamily
  hep-ex/0010015}}].

\bibitem{Fischer:2026bka}
Gfitter Group, Y.~Fischer, J.~Haller, A.~Hoecker, R.~Kogler, F.~Labe,
  K.~M{\"o}nig, D.~Schwarz and J.~Stelzer, \emph{{The Higgs boson through the
  lens of electroweak precision data}},
  \href{https://arxiv.org/abs/2607.09861}{{\ttfamily 2607.09861}}.

\bibitem{ParticleDataGroup:2026}
Particle Data Group, F.~Takahashi et~al., \emph{{Review of particle physics}},
  {\emph{Int. J. Mod. Phys. A} 41 (2026) 2630011}.
  \href{https://pdg.lbl.gov/}{https://pdg.lbl.gov/}.

\bibitem{Brivio:2017btx}
I.~Brivio, Y.~Jiang and M.~Trott, \emph{{The SMEFTsim package, theory and
  tools}}, \href{https://doi.org/10.1007/JHEP12(2017)070}{\emph{JHEP} 12 (2017)
  070}, [\href{https://arxiv.org/abs/1709.06492}{{\ttfamily 1709.06492}}].

\bibitem{Alwall:2014hca}
J.~Alwall et~al., \emph{{The automated computation of tree-level and
  next-to-leading order differential cross sections, and their matching to
  parton shower simulations}},
  \href{https://doi.org/10.1007/JHEP07(2014)079}{\emph{JHEP} 07 (2014) 079},
  [\href{https://arxiv.org/abs/1405.0301}{{\ttfamily 1405.0301}}].

\bibitem{Dawson:2019clf}
S.~Dawson and P.~P. Giardino, \emph{{Electroweak and QCD corrections to $Z$ and
  $W$ pole observables in the Standard Model EFT}},
  \href{https://doi.org/10.1103/PhysRevD.101.013001}{\emph{Phys. Rev. D} 101
  (2020) 013001}, [\href{https://arxiv.org/abs/1909.02000}{{\ttfamily
  1909.02000}}].

\bibitem{Dawson:2022bxd}
S.~Dawson and P.~P. Giardino, \emph{{Flavorful electroweak precision
  observables in the Standard Model effective field theory}},
  \href{https://doi.org/10.1103/PhysRevD.105.073006}{\emph{Phys. Rev. D} 105
  (2022) 073006}, [\href{https://arxiv.org/abs/2201.09887}{{\ttfamily
  2201.09887}}].

\bibitem{Bellafronte:2023amz}
L.~Bellafronte, S.~Dawson and P.~P. Giardino, \emph{{The importance of flavor
  in SMEFT electroweak precision fits}},
  \href{https://doi.org/10.1007/JHEP05(2023)208}{\emph{JHEP} 05 (2023) 208},
  [\href{https://arxiv.org/abs/2304.00029}{{\ttfamily 2304.00029}}].

\bibitem{Dubovyk:2019szj}
I.~Dubovyk, A.~Freitas, J.~Gluza, T.~Riemann and J.~Usovitsch,
  \emph{{Electroweak pseudo-observables and Z-boson form factors at two-loop
  accuracy}}, \href{https://doi.org/10.1007/JHEP08(2019)113}{\emph{JHEP} 08
  (2019) 113}, [\href{https://arxiv.org/abs/1906.08815}{{\ttfamily
  1906.08815}}].

\bibitem{Flacher:2008zq}
{Gfitter Group}, H.~Fl{\"a}cher, M.~Goebel, J.~Haller, A.~Hoecker, K.~M{\"o}nig
  and J.~Stelzer, \emph{{Revisiting the global electroweak fit of the Standard
  Model and beyond with Gfitter}},
  \href{https://doi.org/10.1140/epjc/s10052-009-0966-6,
  10.1140/epjc/s10052-011-1718-y}{\emph{Eur. Phys. J. C} 60 (2009) 543},
  [\href{https://arxiv.org/abs/0811.0009}{{\ttfamily 0811.0009}}]. {Erratum:
  \textit{Eur. Phys. J. C} 71 (2011) 1718}.

\bibitem{Baak:2011ze}
Gfitter Group, M.~Baak, M.~Goebel, J.~Haller, A.~Hoecker, D.~Kennedy,
  K.~Moenig, M.~Schott and J.~Stelzer, \emph{{Updated status of the global
  electroweak fit and constraints on new physics}},
  \href{https://doi.org/10.1140/epjc/s10052-012-2003-4}{\emph{Eur. Phys. J. C}
  72 (2012) 2003}, [\href{https://arxiv.org/abs/1107.0975}{{\ttfamily
  1107.0975}}].

\bibitem{Baak:2012kk}
Gfitter Group, M.~Baak, M.~Goebel, J.~Haller, A.~Hoecker, D.~Kennedy,
  R.~Kogler, K.~Moenig, M.~Schott and J.~Stelzer, \emph{{The electroweak fit of
  the Standard Model after the discovery of a new boson at the LHC}},
  \href{https://doi.org/10.1140/epjc/s10052-012-2205-9}{\emph{Eur. Phys. J. C}
  72 (2012) 2205}, [\href{https://arxiv.org/abs/1209.2716}{{\ttfamily
  1209.2716}}].

\bibitem{Baak:2014ora}
Gfitter Group, M.~Baak, J.~C{\'u}th, J.~Haller, A.~Hoecker, R.~Kogler,
  K.~M{\"o}nig, M.~Schott and J.~Stelzer, \emph{{The global electroweak fit at
  NNLO and prospects for the LHC and ILC}},
  \href{https://doi.org/10.1140/epjc/s10052-014-3046-5}{\emph{Eur. Phys. J. C}
  74 (2014) 3046}, [\href{https://arxiv.org/abs/1407.3792}{{\ttfamily
  1407.3792}}].

\bibitem{Haller:2018nnx}
{Gfitter Group}, J.~Haller, A.~Hoecker, R.~Kogler, K.~M{\"o}nig, T.~Peiffer and
  J.~Stelzer, \emph{{Update of the global electroweak fit and constraints on
  two-Higgs-doublet models}},
  \href{https://doi.org/10.1140/epjc/s10052-018-6131-3}{\emph{Eur. Phys. J. C}
  78 (2018) 675}, [\href{https://arxiv.org/abs/1803.01853}{{\ttfamily
  1803.01853}}].

\bibitem{Wilks:1938dza}
S.~S. Wilks, \emph{{The large-sample distribution of the likelihood ratio for
  testing composite hypotheses}},
  \href{https://doi.org/10.1214/aoms/1177732360}{\emph{Annals Math. Statist.} 9
  (1938) 60}.

\bibitem{Kaplan:1991dc}
D.~B. Kaplan, \emph{{Flavor at SSC energies: a new mechanism for dynamically
  generated fermion masses}},
  \href{https://doi.org/10.1016/S0550-3213(05)80021-5}{\emph{Nucl. Phys. B} 365
  (1991) 259}.

\bibitem{DeSimone:2012fs}
A.~De~Simone, O.~Matsedonskyi, R.~Rattazzi and A.~Wulzer, \emph{{A first top
  partner hunter's guide}},
  \href{https://doi.org/10.1007/JHEP04(2013)004}{\emph{JHEP} 04 (2013) 004},
  [\href{https://arxiv.org/abs/1211.5663}{{\ttfamily 1211.5663}}].

\bibitem{Panico:2015jxa}
G.~Panico and A.~Wulzer, \emph{{The composite Nambu-Goldstone Higgs}},
  vol.~913.
\newblock Springer, 2016,
  \href{https://doi.org/10.1007/978-3-319-22617-0}{10.1007/978-3-319-22617-0}.

\bibitem{Aguilar-Saavedra:2013qpa}
J.~A. Aguilar-Saavedra, R.~Benbrik, S.~Heinemeyer and M.~P{\'e}rez-Victoria,
  \emph{{Handbook of vectorlike quarks: mixing and single production}},
  \href{https://doi.org/10.1103/PhysRevD.88.094010}{\emph{Phys. Rev. D} 88
  (2013) 094010}, [\href{https://arxiv.org/abs/1306.0572}{{\ttfamily
  1306.0572}}].

\bibitem{Agashe:2006at}
K.~Agashe, R.~Contino, L.~Da~Rold and A.~Pomarol, \emph{{A custodial symmetry
  for $Zb \bar b$}},
  \href{https://doi.org/10.1016/j.physletb.2006.08.005}{\emph{Phys. Lett. B}
  641 (2006) 62}, [\href{https://arxiv.org/abs/hep-ph/0605341}{{\ttfamily
  hep-ph/0605341}}].

\bibitem{Contino:2008hi}
R.~Contino and G.~Servant, \emph{{Discovering the top partners at the LHC using
  same-sign dilepton final states}},
  \href{https://doi.org/10.1088/1126-6708/2008/06/026}{\emph{JHEP} 06 (2008)
  026}, [\href{https://arxiv.org/abs/0801.1679}{{\ttfamily 0801.1679}}].

\bibitem{Mrazek:2009yu}
J.~Mrazek and A.~Wulzer, \emph{{A strong sector at the LHC: top partners in
  same-sign dileptons}},
  \href{https://doi.org/10.1103/PhysRevD.81.075006}{\emph{Phys. Rev. D} 81
  (2010) 075006}, [\href{https://arxiv.org/abs/0909.3977}{{\ttfamily
  0909.3977}}].

\bibitem{Matsedonskyi:2012ym}
O.~Matsedonskyi, G.~Panico and A.~Wulzer, \emph{{Light top partners for a light
  composite Higgs}}, \href{https://doi.org/10.1007/JHEP01(2013)164}{\emph{JHEP}
  01 (2013) 164}, [\href{https://arxiv.org/abs/1204.6333}{{\ttfamily
  1204.6333}}].

\bibitem{Azatov:2013hya}
A.~Azatov, M.~Salvarezza, M.~Son and M.~Spannowsky, \emph{{Boosting top partner
  searches in composite Higgs models}},
  \href{https://doi.org/10.1103/PhysRevD.89.075001}{\emph{Phys. Rev. D} 89
  (2014) 075001}, [\href{https://arxiv.org/abs/1308.6601}{{\ttfamily
  1308.6601}}].

\bibitem{Xie:2019gya}
K.-P. Xie, G.~Cacciapaglia and T.~Flacke, \emph{{Exotic decays of top partners
  with charge 5/3: bounds and opportunities}},
  \href{https://doi.org/10.1007/JHEP10(2019)134}{\emph{JHEP} 10 (2019) 134},
  [\href{https://arxiv.org/abs/1907.05894}{{\ttfamily 1907.05894}}].

\bibitem{deBlas:2017xtg}
J.~de~Blas, J.~C. Criado, M.~Perez-Victoria and J.~Santiago, \emph{{Effective
  description of general extensions of the Standard Model: the complete
  tree-level dictionary}},
  \href{https://doi.org/10.1007/JHEP03(2018)109}{\emph{JHEP} 03 (2018) 109},
  [\href{https://arxiv.org/abs/1711.10391}{{\ttfamily 1711.10391}}].

\bibitem{Carmona:2021xtq}
A.~Carmona, A.~Lazopoulos, P.~Olgoso and J.~Santiago, \emph{{Matchmakereft:
  automated tree-level and one-loop matching}},
  \href{https://doi.org/10.21468/SciPostPhys.12.6.198}{\emph{SciPost Phys.} 12
  (2022) 198}, [\href{https://arxiv.org/abs/2112.10787}{{\ttfamily
  2112.10787}}].

\bibitem{Crivellin:2022fdf}
A.~Crivellin, M.~Kirk, T.~Kitahara and F.~Mescia, \emph{{Large
  t{\textrightarrow}cZ as a sign of vectorlike quarks in light of the W mass}},
  \href{https://doi.org/10.1103/PhysRevD.106.L031704}{\emph{Phys. Rev. D} 106
  (2022) L031704}, [\href{https://arxiv.org/abs/2204.05962}{{\ttfamily
  2204.05962}}].

\bibitem{Guedes:2024vuf}
G.~Guedes and P.~Olgoso, \emph{{From the EFT to the UV: the complete SMEFT
  one-loop dictionary}},
  \href{https://doi.org/10.21468/SciPostPhys.20.3.074}{\emph{SciPost Phys.} 20
  (2026) 074}, [\href{https://arxiv.org/abs/2412.14253}{{\ttfamily
  2412.14253}}].

\bibitem{LHCHiggsCrossSectionWorkingGroup:2013rie}
LHC Higgs Cross Section Working Group, J.~R. Andersen et~al., \emph{{Handbook
  of LHC Higgs cross sections: 3. Higgs properties}}.
\newblock CERN Yellow Reports: Monographs. 2013,
  \href{https://doi.org/10.5170/CERN-2013-004}{10.5170/CERN-2013-004}.

\bibitem{CMS:2026nce}
CMS Collaboration, \emph{{Combined measurements and interpretations of Higgs
  boson production and decay in proton-proton collisions at $\sqrt{s}$ = 13
  TeV}},  \href{https://arxiv.org/abs/2602.18611}{{\ttfamily 2602.18611}}.

\bibitem{CMS:2024mke}
CMS Collaboration, \emph{{Measurements of inclusive and differential cross
  sections for top quark production in association with a Z boson in
  proton-proton collisions at $ \sqrt{s} $ = 13 TeV}},
  \href{https://doi.org/10.1007/JHEP02(2025)177}{\emph{JHEP} 02 (2025) 177},
  [\href{https://arxiv.org/abs/2410.23475}{{\ttfamily 2410.23475}}].

\bibitem{ATLAS:2023eld}
ATLAS Collaboration, \emph{{Inclusive and differential cross-section
  measurements of $ t\overline{t}Z $ production in pp collisions at $ \sqrt{s}
  $ = 13 TeV with the ATLAS detector, including EFT and spin-correlation
  interpretations}}, \href{https://doi.org/10.1007/JHEP07(2024)163}{\emph{JHEP}
  07 (2024) 163}, [\href{https://arxiv.org/abs/2312.04450}{{\ttfamily
  2312.04450}}].

\bibitem{ATLAS:2024ojr}
ATLAS Collaboration, \emph{{Measurement of $t$-channel production of single top
  quarks and antiquarks in $pp$ collisions at 13 TeV using the full ATLAS Run 2
  data sample}}, \href{https://doi.org/10.1007/JHEP05(2024)305}{\emph{JHEP} 05
  (2024) 305}, [\href{https://arxiv.org/abs/2403.02126}{{\ttfamily
  2403.02126}}]. [Erratum: JHEP 06 (2025) 024].

\bibitem{ATLAS:2022rms}
ATLAS Collaboration, \emph{{Measurement of the polarisation of $W$ bosons
  produced in top-quark decays using dilepton events at $\sqrt{s} = 13$ TeV
  with the ATLAS experiment}},
  \href{https://doi.org/10.1016/j.physletb.2023.137829}{\emph{Phys. Lett. B}
  843 (2023) 137829}, [\href{https://arxiv.org/abs/2209.14903}{{\ttfamily
  2209.14903}}].

\bibitem{ATLAS:2024vlqpair}
ATLAS Collaboration, \emph{{Search for pair-production of vector-like quarks in
  lepton+jets final states containing at least one $b$-tagged jet using the Run
  2 data from the ATLAS experiment}},
  \href{https://doi.org/10.1016/j.physletb.2024.138743}{\emph{Phys. Lett. B}
  854 (2024) 138743}, [\href{https://arxiv.org/abs/2401.17165}{{\ttfamily
  2401.17165}}].

\bibitem{ATLAS:2026ojf}
ATLAS Collaboration, \emph{{Search for pair-produced vector-like $T$-quarks
  decaying into $Ht$ final states in the lepton-plus-jets channel in $pp$
  collisions at $\sqrt{s}$=13 TeV with the ATLAS detector}},
  \href{https://arxiv.org/abs/2605.28538}{{\ttfamily 2605.28538}}.

\bibitem{ATLAS:2022hnn}
ATLAS Collaboration, \emph{{Search for pair-production of vector-like quarks in
  $pp$ collision events at $\sqrt{s}=13$ TeV with at least one leptonically
  decaying $Z$ boson and a third-generation quark with the ATLAS detector}},
  \href{https://doi.org/10.1016/j.physletb.2023.138019}{\emph{Phys. Lett. B}
  843 (2023) 138019}, [\href{https://arxiv.org/abs/2210.15413}{{\ttfamily
  2210.15413}}].

\bibitem{CMS:2022fck}
CMS Collaboration, \emph{{Search for pair production of vector-like quarks in
  leptonic final states in proton-proton collisions at $\sqrt{s}$ = 13 TeV}},
  \href{https://doi.org/10.1007/JHEP07(2023)020}{\emph{JHEP} 07 (2023) 020},
  [\href{https://arxiv.org/abs/2209.07327}{{\ttfamily 2209.07327}}].

\bibitem{ATLAS:2023singleT}
ATLAS Collaboration, \emph{{Search for single production of vector-like $T$
  quarks decaying into $Ht$ or $Zt$ in $pp$ collisions at $\sqrt{s}=13$ TeV
  with the ATLAS detector}},
  \href{https://doi.org/10.1007/JHEP08(2023)153}{\emph{JHEP} 08 (2023) 153},
  [\href{https://arxiv.org/abs/2305.03401}{{\ttfamily 2305.03401}}].

\bibitem{ATLAS:2025singleWb}
ATLAS Collaboration, \emph{{Search for single production of vector-like quarks
  decaying into $W(\ell\nu)b$ in $pp$ collisions at $\sqrt{s}=13$ TeV with the
  ATLAS detector}}, \href{https://doi.org/10.1007/JHEP12(2025)012}{\emph{JHEP}
  12 (2025) 012}, [\href{https://arxiv.org/abs/2506.15515}{{\ttfamily
  2506.15515}}].

\bibitem{CMS:2026singleWb}
CMS Collaboration, \emph{{Search for the single production of vector-like
  quarks decaying into a $W$ boson and a $b$ quark using single-lepton final
  states in proton-proton collisions at $\sqrt{s}=13$ TeV}},
  \href{https://arxiv.org/abs/2604.17564}{{\ttfamily 2604.17564}}. Submitted to
  Phys. Lett. B.

\bibitem{DAmbrosio:2002vsn}
G.~D'Ambrosio, G.~F. Giudice, G.~Isidori and A.~Strumia, \emph{{Minimal flavor
  violation: an effective field theory approach}},
  \href{https://doi.org/10.1016/S0550-3213(02)00836-2}{\emph{Nucl. Phys. B} 645
  (2002) 155}, [\href{https://arxiv.org/abs/hep-ph/0207036}{{\ttfamily
  hep-ph/0207036}}].

\bibitem{Barbieri:2012uh}
R.~Barbieri, D.~Buttazzo, F.~Sala and D.~M. Straub, \emph{{Flavour physics from
  an approximate $U(2)^3$ symmetry}},
  \href{https://doi.org/10.1007/JHEP07(2012)181}{\emph{JHEP} 07 (2012) 181},
  [\href{https://arxiv.org/abs/1203.4218}{{\ttfamily 1203.4218}}].

\end{thebibliography}\endgroup

\end{document}